\documentclass[onecolumn,12pt,prb]{revtex4}%
\usepackage{makeidx}
\usepackage{amsmath}
\usepackage{graphicx}
\usepackage{dcolumn}
\usepackage{bm}
\usepackage{color}%
\usepackage{amsfonts}%
\usepackage{amssymb}
\begin{document}

\title{The quantum $J_{1}-J_{1}^{\prime}-J_{2}$ spin-$1/2$ Heisenberg
antiferromagnet: A variational method \ study}
\author{Orlando D. Mabelini $^{1}$, Octavio Salmon$^{1},$and J. Ricardo de
Sousa$^{1,2}$\\$^{1}$Universidade Federal do Amazonas, Departamento de F\'{\i}sica, \\3000, Japiim, \ 69077-000, Manaus-AM, Brazil\\$^{2}$National Institute of Science and Technology for Complex Systems\\Universidade Federal do Amazonas, Departamento de F\'{\i}sica, \\3000, Japiim, 69077-000, Manaus-AM, Brazil }

\begin{abstract}
The phase transition of the quantum spin-$1/2$ frustrated Heisenberg
antiferroferromagnet on an anisotropic square lattice is studied by using a
variational treatment. The model is described by the Heisenberg Hamiltonian
with two antiferromagnetic interactions: nearest-neighbor (NN) with different
coupling strengths $J_{1}$ and $J_{1}^{\prime}$ along x and y directions
competing with a next-nearest-neighbor coupling $J_{2}$ (NNN). The ground
state phase diagram in the ($\lambda,\alpha$) space, where $\lambda
=J_{1}^{\prime}/J_{1}$ and $\alpha=J_{2}/J_{1}$, is obtained. Depending on the
values of $\lambda$ and $\alpha$, we obtain three different states:
antiferromagnetic (\textbf{AF}), collinear antiferromagnetic (\textbf{CAF})
and quantum paramagnetic (\textbf{QP}). For an intermediate region
$\lambda_{1}<\lambda<1$ we observe a \textbf{QP} state between the ordered
\textbf{AF} and \textbf{CAF} phases, which disappears for $\lambda$ above some
critical value $\lambda_{1}$. The boundaries between these ordered phases
merge at the \textit{quantum critical endpoint} (\textbf{QCE}). Below this
\textbf{QCE} there is again a direct first-order transition between the
\textbf{AF} and \textbf{CAF} phases, with a behavior approximately described
by the classical line $\alpha_{c}\simeq\lambda/2$.


\end{abstract}
\maketitle

\section{Introduction}

The study of the phase transition of frustrated spin systems on
two-dimensional (2d) lattices is a central problem in modern condensed mater
physics. A competition of exchange interaction can lead to frustration, where
spatial arrangement of magnetic ions in a crystal for which a simultaneous
antiparallel ordering of all interacting spin is impossible. In particular,
one of the frustrated 2d models most discussed is the quantum spin-$1/2$
Heisenberg antiferromagnet on a square lattice with competing nearest-neighbor
(NN) and next-nearest-neighbor (NNN) antiferromagnetic exchange interactions
(known as $J_{1}-J_{2}$ model) \cite{1,3,4,5,6,7,8, 9,
10,11,12,bishop1,darradi,isaev,viana,oliveira}.

The criticality of this $J_{1}-J_{2}$ Heisenberg model on a square lattice are
relatively well known at $T=0$. \ There are two magnetically long-range
ordered phases at small and at large values of $\alpha=J_{2}/J_{1}$ separated
by an intermediate quantum paramagnetic phase without magnetic long-range
order in the region between $\alpha_{1c}\simeq0.4$ and $\alpha_{2c}\simeq0.6$,
where the properties of these disordered phase are still under intensive
debate. For $\alpha<$ $\alpha_{1c}$, the system possesses antiferromagnetic
(AF) long-range order with wave vector $\mathbf{Q}=(\pi,\pi)$, with a
staggered magnetization smaller than the saturated value (quantum
fluctuations), which vanished continuously when $\alpha\rightarrow\alpha_{1c}%
$. For $\alpha>\alpha_{2c}$ we have two degenerate collinear states which are
the helical states with pitch vectors $\mathbf{Q}=(\pi,0)$ and $(0,\pi)$.
These two collinear states are characterized by a parallel spin orientation of
nearest neighbors in vertical (or horizontal) direction and an antiparallel
spin orientation of nearest neighbors in horizontal (or vertical) direction,
and therefore exhibit N\'{e}el order within the initial sublattice A and B. At
$\alpha=\alpha_{2c}$, the magnetization jumps from a nonzero to a zero value.
The phase transition from N\'{e}el to the quantum paramagnetic state is second
order, whereas the transition from the collinear to the quantum paramagnetic
state is first order\cite{viana,oliveira}. Isaev, \textit{et al}.\cite{isaev}
have shown that the intermediate quantum paramagnetic is a (\textit{singlet})
plaquette crystal, and the ground and first excited states are separated by a
finite gap.

The interest to study the two-dimensional $J_{1}-J_{2}$ Heisenberg
antiferromagnet have been greatly stimulated by its experimental realization
in vanadium phosphates compounds\cite{melzi, carretta, rosner, bombardi}, such
as Li$_{2}$VOSiO$_{4}$, Li$_{2}$VOGeO$_{4}$, and \ VOMoO$_{4}$, which might be
described by this frustrated model in the case of $J_{2}\simeq J_{1}$
($\alpha=J_{2}/J_{1}\simeq1$). These isostructural compounds are characterized
by a layered structure containing V$^{4+}$ ($S=1/2$) ions. The structure of
V$^{4+}$ layer suggest that the superexchange is similar. In these compounds a
second order phase transition to a long-range ordered magnetic phase has been
observed. NMR spin-lattice relaxation measurements\cite{melzi} below $T_{c}$
shows that the order is collinear. Due to the two-fold degeneracy of the
ground-state for $\alpha>0.7$ it is not possible to say \textit{a priori}
which will be the magnetic wave vector (i.e., $\mathbf{Q}=(\pi,0)$ and
$(0,\pi)$) below $T_{c}$. On the other hand, such a scenario can change by
considering spin-lattice coupling which will lift the degeneracy of the
ground-state and will lower its energy\cite{becca}. Then, any structural
distortion should inevitably reduce this competing interactions and thus
reduces the frustration. In the case of \ this frustrated magnetic materials,
the competing interactions are inequivalent but their topology and magnitudes
can be tuned so that the strong quantum fluctuations destroy the long-range
ordering. Experimentally the ground state phase diagram of frustrated
compounds, described by the $J_{1}-J_{2}$ model, can be explored continuously
from high to the low $\alpha=J_{2}/J_{1}$ regime by applying high pressures
(P), which modify the bonding lengths and angles. Recent results from x-ray
diffraction measurements\cite{pavarini} on the Li$_{2}$VOSiO$_{4}$ compound
has shown that the ratio $\alpha$ decreases by about $40\%$ when the pressure
increases from $0$ to $7.6$GPa. \ 

A generalization of the $J_{1}-J_{2}$ Heisenberg antiferromagnetic model on a
square lattice was introduced by Nersesyan and Tsvelik\cite{nerseyan} and
studied by other groups\cite{equivalence1, equivalence2, equivalence3,
equivalence4, equivalence5,equivalence6,equivalence7,equivalence8}, the
so-called $J_{1}-J_{1}^{\prime}-J_{2}$ model. In the $J_{1}-J_{1}^{\prime
}-J_{2}$ model is considered inequivalence nn couplings $J_{1}$ and
$J_{1}^{\prime}=\lambda J_{1}$ in the two orthogonal spatial lattice
dimensions with all the NNN bonds across the diagonals to have the same
strength $J_{2}$. Study of extensive band structure
calculations\cite{equivalence6} for the vanadium phosphates ABVO(PO$_{4}%
$)$_{2}$ (AB=Pb$_{2}$, SrZn, BaZn, and BaCd) have indicated four inequivalent
exchange couplings: $J_{1}$ and $J_{1}^{\prime}$ between NN and $J_{2}$ and
$J_{2}^{\prime}$ between NNN. For example, in SrZnVO(PO$_{4}$)$_{2}$ was
estimated $J_{1}^{\prime}/J_{1}\simeq0.7$ and $J_{2}^{\prime}/J_{2}\simeq0.4$
causing a distortion of the spin lattice. This spatial anisotropy tends to
narrow the critical region and destroys it completely at a certain value of
the interchain parameter $\lambda$.

On the other hand, by using the continuum limit of the $J_{1}-J_{1}^{\prime
}-J_{2}$ spin-$1/2$ model Starykh and Balents\cite{equivalence1} have shown
that this transition splits into two, with the presence of an intermediate
quantum paramagnetic (columnar dimer) phase for $\lambda\leq1$. Bishop,
\textit{et al}\cite{equivalence4}, by using coupled cluster treatment found
the surprising and novel result that there exists a quantum triple point
(\textbf{QTP}) with coordinates at ($\alpha_{t}=0.33\pm0.02,\lambda
_{t}=0.60\pm0.03$), below which there is a second-order phase transition
between the \textbf{AF} and \textbf{CAF} phases while above this \textbf{QTP}
are these two ordered phases separated by the intermediate magnetically
disordered phase (VBS or RVB). The order parameters of both the \textbf{AF}
and \textbf{CAF} phases vanish continuously both below and above the
\textbf{QTP}, which is typical of second-order phase transition. There is some
evidence that the transition between the \textbf{CAF} and intermediate phases
is of first-order. Using exact diagonalization\cite{equivalence2} with small
lattice of $N\leq36$ ($6\times6$) size, the intermediate \textbf{QP} phase for
all interval of $\lambda\in\lbrack0,1]$ has been obtained for the pure
spin-$1/2$ $J_{1}-J_{2}$ model on a square lattice. These results are in
accordance with results obtained by Starykh and Balentes\cite{equivalence1},
that predicted not the \textbf{QTP} in the ground-state phase diagram recently
observed by Bishop, \textit{et al.}\cite{equivalence4}.

The ground state (GS) properties of the two-dimensional frustrated Heisenberg
antiferromagnet have been investigated by various methods. The exact
diagonalization starts from singlet states on pairs of sites, which cover the
whole 2d lattice. However, the manifold of these states which can be
constructed is nonorthogonal and overcomplete. This numerical methods are
limited to small clusters $N\leq6\times6=36$ due to storage problems. The
computation on the largest cluster $6\times6$ has been performed by Schulz and
co-workes\cite{3} $20$ years ago. In spite of the great improvements achieved
during this time, it is not possible so far to repeat this calculation for the
next interesting cluster $8\times8$. This is only possible with other
technique, as the quantum Monte Carlo simulation. Due to the progress in
computer hardware and the increased efficiency in programing, very
recently\cite{exact} the GS of the quantum spin-1/2 $J_{1}-J_{2}$ model have
been calculated by the Lanczos algoritm for a square lattice with $N=40$ \ sites.

The theoretical treatment of the frustrated quantum models is far from being
trivial. Many of the standard many-body methods, such as quantum Monte Carlo
techniques, may fail or become computationally infeasible to implement if
frustration is present due to the minus-sign problem. Hence, there is
considerable interest in any method that can deal with frustrated spin
systems. This considerable qualitative difference in the ground state phase
diagram in the $\alpha-\lambda$ plane of the quantum spin-$1/2$ $J_{1}%
-J_{1}^{\prime}-J_{2}$ model further motivates us to study this issue by
alternative methods.

Using a variational approximation, in which plaquettes of four spins are
treated exactly, Oliveira\cite{oliveira} has studied the ground state phase
diagram of the pure $J_{1}-J_{2}$ Heisenberg antiferromagnet on a square
lattice, where the quantitative results are in good accordance with a more
sophisticated method (exact diagonalization). In this work, we generalize this
variational method to treat the anisotropic square lattice ( $J_{1}%
-J_{1}^{\prime}-J_{2}$ model). The rest of this paper is organized as follows:
In Sec. II, the model is presented and a brief discussion of results. In Sec.
III, the method is applied for the case of one plaquette with four spins
interacting with other plaquette type mean field approximation. Main results
will be presented in Sec. IV, as well as some discussions. Finally, in Sec. V
we will give a brief summary.

\section{Model}

The critical behavior of the quantum spin-$1/2$ $J_{1}-J_{2}$ Heisenberg model
has been studied for many years, but very little has been done in the
anisotropic square lattice case, which is described by following Hamiltonian:%

\begin{equation}
\mathcal{H}=%
%TCIMACRO{\tsum \limits_{\left\langle i,j\right\rangle }}%
%BeginExpansion
{\textstyle\sum\limits_{\left\langle i,j\right\rangle }}
%EndExpansion
\left(  J_{1}\mathbf{\sigma}_{i,j}\cdot\mathbf{\sigma}_{i+1,j}+J_{1}^{\prime
}\mathbf{\sigma}_{i,j}\cdot\mathbf{\sigma}_{i,j+1}\right)  +J_{2}%
%TCIMACRO{\tsum \limits_{\left\langle \left\langle i,j\right\rangle
%\right\rangle }}%
%BeginExpansion
{\textstyle\sum\limits_{\left\langle \left\langle i,j\right\rangle
\right\rangle }}
%EndExpansion
\left(  \mathbf{\sigma}_{i,j}\cdot\mathbf{\sigma}_{i+1,j+1}+\mathbf{\sigma
}_{i+1,j}\cdot\mathbf{\sigma}_{i,j+1}\right)  , \tag{1}%
\end{equation}
where $\mathbf{\sigma}_{i,j}\mathbf{=}(\sigma_{i,j}^{x},\sigma_{i,j}%
^{y},\sigma_{i,j}^{z})$ is the spin-$1/2$ Pauli spin operators, the index
$(i,j)$ labels the $x$ (row) and $y$ (column) components of the lattice sites.
The first sum runs over all NN and the second sum runs over all NNN pairs. We
denote the Hamiltonian (1) \ by $J_{1}-J_{1}^{\prime}-J_{2}$ model, with
strength $J_{1}$ along the row direction, $J_{1}^{\prime}=\lambda J_{1}$ along
the column direction, $J_{2}=\alpha J_{1}$ along the diagonals, and we assume
all couplings to be positive with $J_{1}^{\prime}<J_{1}$.

The classical ($S=\infty$) model (1) has only two ordered ground-states:
\textbf{AF} (or N\'{e}el) for $\alpha>\lambda/2$ and columnar stripe
(\textbf{CAF}) for $\alpha<\lambda/2$ separated by a first-order line at
$\alpha_{c}=\lambda/2$. Quantum fluctuations play a significant role in the
magnetic phase diagram of the system at zero temperature. We will investigate
the role of quantum fluctuations on the stability of the N\'{e}el and
collinear phases. In the $S=1/2$ case (quantum limit), the line splits into
two phase transitions, where the ordered states (\textbf{AF} and \textbf{CAF})
are separated by an intermediate quantum paramagnetic (\textbf{QP}) phase,
both on a square lattice. Exact diagonalization\cite{12} has estimated a
critical line at $\alpha_{CAF}=\alpha_{c}+\frac{5\lambda^{2}}{8\pi^{2}}$, for
the transition between the \textbf{CAF} and \textbf{QP} states, and at
$\alpha_{AF}=\alpha_{c}-\frac{\lambda^{2}}{8\pi^{2}}$ between the \textbf{AF}
and \textbf{QP} states. The phase diagram in the $\alpha-\lambda$ plane
obtained is in accordance with Starykh and Balents\cite{equivalence1}.
However, the existence of \textbf{QTP} (\textit{quantum triple point}) that
was predicted by Bishop, \textit{et al}.\cite{equivalence4}, is not present in
their obtained phase diagram. Moreover, they found only presence of
second-order phase transitions in the phase diagram. This contradictory
qualitative results (existence or not of \textbf{QTP}) is the primary
motivation behind this present work. 

On the other hand, a critical endpoint (\textbf{CE}) is a point in the phase
diagram where a critical line meets and is truncated by a first-order line.
This \textbf{CE} appear in the phase diagram of many physical systems such as
binary fluid mixtures, superfluids, binary alloys, liquid crystals, certain
ferromagnets, etc, and have been known for over a century\cite{ce1}. Despite
the \textbf{CE} long history, new singularities at the \textbf{CE} were
predicted. Fisher and Upton\cite{ce2} argued that a new singularity in the
curvature of the first-order phase transition line should arise at a
\textbf{CE}. This prediction was confirmed by Fisher and Barbosa's\cite{ce3}
phenomenological studies for an exactly solvable spherical model. In
conclusion of the analysis of the multicritical behavior observed in the
ground-state phase diagram in the $\lambda-\alpha$ plane for the $J_{1}%
-J_{1}^{\prime}-J_{2}$ model, we have the presence of a \textit{quantum
critical endpoint} (\textbf{QCE}) and not \textbf{QTP} as mentioned other
works\cite{equivalence4,equivalence8} Therefore,  the objective of this work
is to obtain the \textbf{QCE} using the variational method, that was developed
previously by Oliveira\cite{oliveira} in the pure limit ($\lambda=1$) case.

\section{Method}

We first express the fluctuations around the classical ground state
(\textbf{AF} and \textbf{CAF} phases), where consider a trial vector state
$\left|  \Psi_{0}\right\rangle =\prod\limits_{l=1}^{N/4}\left|  \phi
_{0l}\right\rangle $ for the ground state as a product of plaquette state
$\left|  \phi_{0l}\right\rangle $ . We denote the plaquettes by $l$ label,
that is composed of four spins, where it do not overlap (\textit{mean field})
on the square lattice as illustrated in figure 1. Each plaquette state is
given by%

%TCIMACRO{\FRAME{ftbpFU}{309.6875pt}{266.75pt}{0pt}{\Qcb{Two-dimensional square
%lattice with a plaquette structure to be considered in this paper. The negrit
%plaquette is composed with the four $\mathbf{\sigma}_{1},\mathbf{\sigma}%
%_{2},\mathbf{\sigma}_{3},$ and $\mathbf{\sigma}_{4\text{ }}$ spin operators
%that are considered in Eq. (2).}}{}{Figure}%
%maintain-aspect-ratio TRUE;  display "USEDEF";  valid_file "T";
%width 309.6875pt;  height 266.75pt;  depth 0pt;  original-width 432.3125pt;
%original-height 371.9375pt;  cropleft "0";  croptop "1";  cropright "1";
%cropbottom "0";  tempfilename 'M3HXYI00.wmf';tempfile-properties "XPR";}}}%
%BeginExpansion
\begin{figure}[ptb]
\begin{center}
\includegraphics[natheight=371.937500pt,natwidth=432.312500pt,height=266.75pt,width=309.6875pt]{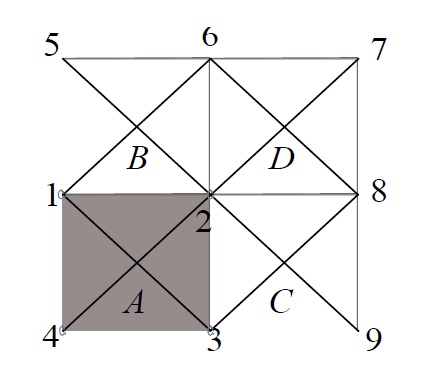}
\caption{Two-dimensional square lattice with a plaquette structure to be
considered in this paper. The negrit plaquette is composed with the four
$\mathbf{\sigma}_{1},\mathbf{\sigma}_{2},\mathbf{\sigma}_{3},$ and
$\mathbf{\sigma}_{4\text{ }}$ spin operators that are considered in Eq. (2).}
\end{center}
\end{figure}
%EndExpansion%

\begin{equation}
\left|  \phi_{0l}\right\rangle =\sum_{n=1}^{6}a_{n}\left|  n\right\rangle
_{l}, \tag{2}%
\end{equation}
where \{$\left|  1\right\rangle =\left|  +-+-\right\rangle $, $\left|
2\right\rangle =\left|  -+-+\right\rangle ,\left|  3\right\rangle =\left|
++--\right\rangle ,\left|  4\right\rangle =\left|  -++-\right\rangle ,\left|
5\right\rangle =\left|  --++\right\rangle ,\left|  6\right\rangle =\left|
+--+\right\rangle $\} is the vector basis with $\sigma^{z}=\sum\limits_{i=1}%
^{4}\sigma_{i}^{z}=0$, \{$a_{i}$\} are real variational parameters obeying the
normalization condition $\sum\limits_{n=1}^{6}a_{n}^{2}=1$. With this choice
of vector states, the mean value of the spin operator in each site of the
plaquette is given by $\left\langle \mathbf{\sigma}_{i}\right\rangle
=m_{i}\widehat{z},$ $m_{i}=\left\langle \sigma_{i}^{z}\right\rangle
=\left\langle \phi_{0l}\left|  \sigma_{i}^{z}\right|  \phi_{0l}\right\rangle
$, where the components in the $x$ and $y$ directions are null.

Using the trial vector state defined in the Eq. (2), we obtain the
magnetizations at each site that are given by%

\begin{equation}
m_{1}=2\left(  xu+yv-zw\right)  , \tag{3}%
\end{equation}%
\begin{equation}
m_{2}=2\left(  -xu+yv+zw\right)  , \tag{4}%
\end{equation}%
\begin{equation}
m_{3}=2\left(  xu-yv+zw\right)  , \tag{5}%
\end{equation}
and%
\begin{equation}
m_{4}=2\left(  -xu-yv+zw\right)  , \tag{6}%
\end{equation}
where we have used the same set of parameters (\textit{canonical
transformation}) of Ref.\cite{oliveira}, i.e., $x=\left(  a_{1}+a_{2}\right)
/\sqrt{2},y=\left(  a_{3}+a_{5}\right)  /\sqrt{2},z=\left(  a_{4}%
+a_{6}\right)  /\sqrt{2},u=\left(  a_{1}-a_{2}\right)  /\sqrt{2},v=\left(
a_{3}-a_{5}\right)  /\sqrt{2}$, and $w=\left(  a_{4}-a_{6}\right)  /\sqrt{2}$,
which obeys the normalization condition $x^{2}+y^{2}+z^{2}+u^{2}+v^{2}%
+w^{2}=1$.

The ground state energy per spin and unit of $J_{1}$, $E_{0}=\left\langle
\Psi_{0}\left|  \mathcal{H}\right|  \Psi_{0}\right\rangle /NJ_{1}$, is given by%

\begin{equation}
E_{0}=E_{01}+E_{02}, \tag{7}%
\end{equation}
with%

\begin{align}
E_{01}  &  =\left\langle \overrightarrow{\sigma}_{1}.\overrightarrow{\sigma
}_{2}\right\rangle _{A}+\lambda\left\langle \overrightarrow{\sigma}%
_{2}.\overrightarrow{\sigma}_{3}\right\rangle _{A}+\left\langle
\overrightarrow{\sigma}_{3}.\overrightarrow{\sigma}_{4}\right\rangle
_{A}+\lambda\left\langle \overrightarrow{\sigma}_{4}.\overrightarrow{\sigma
}_{1}\right\rangle _{A}+\nonumber\\
&  \lambda\left\langle \overrightarrow{\sigma}_{1}\right\rangle _{A}%
.\left\langle \overrightarrow{\sigma}_{5}\right\rangle _{B}+\lambda
\left\langle \overrightarrow{\sigma}_{2}\right\rangle _{A}.\left\langle
\overrightarrow{\sigma}_{6}\right\rangle _{B}+\left\langle \overrightarrow
{\sigma}_{2}\right\rangle _{A}.\left\langle \overrightarrow{\sigma}%
_{8}\right\rangle _{D}+\left\langle \overrightarrow{\sigma}_{3}\right\rangle
_{A}.\left\langle \overrightarrow{\sigma}_{9}\right\rangle _{D}, \tag{8}%
\end{align}
and%
\begin{align}
E_{02}  &  =\alpha\left[  \left\langle \overrightarrow{\sigma}_{1}%
.\overrightarrow{\sigma}_{3}\right\rangle _{A}+\left\langle \overrightarrow
{\sigma}_{2}.\overrightarrow{\sigma}_{4}\right\rangle _{A}+\left\langle
\overrightarrow{\sigma}_{1}\right\rangle _{A}.\left\langle \overrightarrow
{\sigma}_{6}\right\rangle _{B}+\left\langle \overrightarrow{\sigma}%
_{2}\right\rangle _{A}.\left\langle \overrightarrow{\sigma}_{5}\right\rangle
_{B}+\left\langle \overrightarrow{\sigma}_{2}\right\rangle _{A}.\left\langle
\overrightarrow{\sigma}_{7}\right\rangle _{C}+\right. \nonumber\\
&  \left.  \left\langle \overrightarrow{\sigma}_{2}\right\rangle
_{A}.\left\langle \overrightarrow{\sigma}_{9}\right\rangle _{D}+\left\langle
\overrightarrow{\sigma}_{3}\right\rangle _{A}.\left\langle \overrightarrow
{\sigma}_{8}\right\rangle _{D}+\left\langle \overrightarrow{\sigma}%
_{6}\right\rangle _{B}.\left\langle \overrightarrow{\sigma}_{8}\right\rangle
_{D}\right]  , \tag{9}%
\end{align}
where $\left\langle \mathcal{O}\right\rangle _{\mu}=\left\langle \phi_{0\mu
}\left|  \mathcal{O}\right|  \phi_{0\mu}\right\rangle $ is the mean value of a
given observable $\mathcal{O}$ calculated in the vector state of the
$\mu(=A,B,C,D)$ plaquette as illustrated in Fig. 1.

The variational energy can be evaluated using the properties of the spin-$1/2$
Pauli operator components, i.e., $\sigma^{z}\left|  \pm\right\rangle
=\pm\left|  \pm\right\rangle $,$\sigma^{x}\left|  \pm\right\rangle =\left|
\mp\text{ }\right\rangle $ and $\sigma^{y}\left|  \pm\right\rangle =\pm
i\left|  \mp\right\rangle $, \ that is expressed for%
\begin{align}
E_{0}  &  =-\left(  \dfrac{\lambda+1}{2}\right)  \left(  x^{2}+u^{2}\right)
-2\left(  \lambda+1\right)  x^{2}u^{2}+\frac{\left(  1-\lambda\right)  }%
{2}\left[  \left(  y^{2}+v^{2}\right)  -\left(  z^{2}+\omega^{2}\right)
^{2}\right]  +\nonumber\\
&  2\left(  1-\lambda\right)  \left(  y^{2}v^{2}-z^{2}\omega^{2}\right)
+2x\left(  y\lambda+z\right)  +\alpha\left[  \dfrac{1}{2}-\left(  y-z\right)
^{2}-v^{2}-\omega^{2}+6x^{2}u^{2}-6y^{2}v^{2}-6z^{2}\omega^{2}\right]  .
\tag{10}%
\end{align}

To obtain the minimum energy with a boundary condition given by normalization
$x^{2}+y^{2}+z^{2}+u^{2}+v^{2}+w^{2}=1$, we use the Lagrange multiplier method
which correspond the minimization of the functional%
\begin{equation}
\mathcal{F}(x,y,z,u,v,w,\eta)=E_{0}-\eta\left(  x^{2}+y^{2}+z^{2}+u^{2}%
+v^{2}+w^{2}-1\right)  . \tag{11}%
\end{equation}

The stationary solutions ($\delta\mathcal{F}=0$) are obtained by solving the
set of nonlinear equations
\begin{equation}
\left\{
\begin{array}
[c]{c}%
-\left(  \lambda+1\right)  x-4\left(  \lambda+1\right)  xu^{2}+2\left(
y\lambda+z\right)  +12\alpha xu^{2}=2\eta x\\
\left(  1-\lambda\right)  \left(  y+4yv^{2}\right)  +2x\lambda-2\alpha\left(
y-z\right)  -12\alpha yv^{2}=2\eta y\\
-\left(  1-\lambda\right)  z-4\left(  1-\lambda\right)  z\omega^{2}%
+2x+2\alpha\left(  y-z\right)  -12\alpha z\omega^{2}=2\eta z\\
-\left(  \lambda+1\right)  u=2\eta u\\
\left(  1-\lambda\right)  v-2\alpha v=2\eta v\\
-\left(  1-\lambda\right)  \omega-2\alpha\omega=2\eta\omega
\end{array}
\right.  , \tag{12}%
\end{equation}
where $\eta$ is the Lagrange multiplier.

\section{Results}

The variational parameters $x,y,z,u,v,w$, and $\eta$ are determined
simultaneously solving the system of equations (12) combined with the
normalization condition $x^{2}+y^{2}+z^{2}+u^{2}+v^{2}+w^{2}=1$ for each
phase. In the quantum paramagnetic (\textbf{QP}) phase we have $m_{1}%
=m_{2}=m_{3}=m_{4}=0$. We note that in the isotropic limit ($\lambda=1$), our
results reduce the same expression obtained by Oliveira\cite{oliveira}. In
this disordered phase, the ground state vector $\left|  \phi_{0l}\right\rangle
$ is an eigenvector of $\mathbf{S}_{l}^{2}$, where $\mathbf{S}_{l}$ is the
total spin of the $l$th plaquette of four spins, with zero eigenvalue (singlet
state). In the \textbf{AF} ordered phase we have the boundary condition
$m_{1}=-m_{2}=m_{3}=-m_{4}$, and in the \textbf{CAF} phase $m_{1}=m_{2}%
=-m_{3}=-m_{4}$.

The order parameters $m_{AF}=(m_{1}-m_{2}+m_{3}-m_{4})/4$ and $m_{CAF}%
=(m_{1}+m_{2}-m_{3}-m_{4})/4$ are numerically obtained as a function of
frustration parameter $\alpha$ for a given value of spatial anisotropy
$\lambda$. We observe that the order parameter $m_{AF}$ goes smoothly to zero
when the frustration parameter ($\alpha$) increases from zero to $\alpha
_{1c}(\lambda)$ with $\lambda>\lambda_{1}$ characterizing a second-order phase
transition. A simple fitting of the form $m_{AF}\simeq(\alpha_{1c}%
-\alpha)^{\beta}$ in the vicinity of the second-order transition gives the
same classical value for the critical exponent $\beta=1/2$. On the other hand,
for $\lambda>\lambda_{1}$ and $\alpha>\alpha_{2c}(\lambda)$ the staggered
magnetization $m_{CAF}$ increases monotonically with the frustration parameter
$\alpha$ in the \textbf{CAF} phase, with a discontinuity of $m_{CAF}$ at
$\alpha=\alpha_{2c}(\lambda)$, which is a first-order phase transition. For
$\lambda<\lambda_{1}$, the \textbf{QP} intermediate phase between the two
ordered states (\textbf{AF} and \textbf{CAF}) disappears, and a direct
transition between the magnetically ordered \textbf{AF} and \textbf{CAF}
located at the crossing point $\alpha_{c}\simeq\lambda/2$ correspondent to the
classical value.

The ground state ($T=0$) phase diagram in the $\lambda-\alpha$ plane is
displayed in Fig. 2. The solid line indicate the critical points and the
dashed lines represent first-order frontiers. We observe three different
phases, namely: \textbf{AF} (antiferromagnetic), \ \textbf{CAF} (collinear
antiferromagnetic) and \textbf{QP} (quantum paramagnetic). The \textbf{AF} and
\textbf{QP} phases are separated by a second-order transition line
$\alpha_{1c}(\lambda)$, while the \textbf{QP} and \textbf{CAF} phases are
separated by a first-order transition line $\alpha_{2c}(\lambda)$. The
presence of the interchain parameter $\lambda$ has the general effect of
suppressing the \textbf{QP} phase. The \textbf{QP} region decreases gradually
with the decrease of the $\lambda$ parameter, and it disappears completely at
the \textit{quantum} \textit{critical endpoint} \textbf{QCE}$\equiv$%
($\lambda_{1},\alpha_{1}$) where the boundaries between these phases emerge.
Below this \textbf{QCE}, i.e., for $\lambda<\lambda_{1}$, there is a direct
first-order phase transition between the \textbf{AF} and \textbf{CAF }phases,
with a transition point $\alpha_{c}\approx\lambda/2$ (classical value).

In order to illustrate the nature of the phase transition, we also show, in
inset Fig. 2, the behavior of the staggered magnetization (order parameter) as
a function of the frustration parameter ($\alpha$) for $\lambda=0.4(<\lambda
_{1})$ and $0.80$ $(>\lambda_{1})$. From curves such as those shown in Fig. 2
\ we see that for $\lambda=0.8$ there exists an intermediate region between
the critical point $\alpha=\alpha_{1c}(\lambda)$ at which $m_{AF}%
(\alpha)\rightarrow0$ for the \textbf{AF} phase, characterizing a second-order
transition, and the point $\alpha=\alpha_{2c}^{\ast}(\lambda)$ at which the
$m_{CAF}(\alpha)$ order parameter presents a discontinuity for the
\textbf{CAF} phase, characterizing a first-order transition. For $\lambda
=0.4$, the order parameter of the \textbf{AF} phase decreases monotonically
with increase of the frustration parameter from $\approx0.81$, for $\alpha=0$,
to zero for $\alpha\simeq0.2$ ($\simeq\lambda/2$). In the \textbf{CAF} phase
$m_{CAF}(\alpha)$ decreases from $\approx0.95$ for $\alpha=1.0$ to
$\approx0.82$ for $\alpha=\alpha_{2c}^{\ast}(\lambda)\simeq\lambda/2$,
characterizing a direct first-order transition between the magnetically
ordered \textbf{AF} and \textbf{CAF} phases located at the crossing point. We
note that the definition of the order parameter $m_{\mu}=\left\langle
\sigma_{i}^{z}\right\rangle =\left\langle \phi_{0l}\left|  \sigma_{i}%
^{z}\right|  \phi_{0l}\right\rangle $ ($\mu=AF,CAF$) difer of $1/2$ factor
when compared with calculations which use other methods (i.e., $m_{\mu
}=\left\langle S_{i}^{z}\right\rangle =\left\langle \phi_{0l}\left|
\sigma_{i}^{z}\right|  \phi_{0l}\right\rangle /2$). Therefore, in the limit of
the not frustrated ($\alpha=0$) square lattice ($\lambda=1$)
antiferromagnetic, solving the equations (12) and applying the corrections
factor we found $m_{AF}=0.41$ which is consistent with the numerical results
obtained by various methods such as series expansion, quantum Monte Carlo
simulation, and others\cite{heisenberg}, and can also be compared with
experimental results for the K$_{2}$NiF$_{4}$, K$_{2}$MnF$_{4}$, and Rb$_{2}%
$MnF$_{4}$ compounds\cite{24,25,26}.%

%TCIMACRO{\FRAME{ftbpFU}{424.125pt}{374.4375pt}{0pt}{\Qcb{Ground state phase
%diagram in the $\lambda-\alpha$ plane for the quantum spin-$1/2$ $J_{1}%
%-J_{1}^{\prime}-J_{2}$ model on a square lattice, where $\alpha=J_{2}/J_{1}$
%and $\lambda=J_{1}^{\prime}/J_{1}$. The dashed and solid lines corresponds the
%first- and second-order transitions lines, respectively. The black point
%represents the quantum critical endpoint (\textbf{QCE}). The notations
%indicated by \textbf{AF}, \textbf{CAF} and \textbf{QP} corresponds the
%antiferromagnetic, collinear antiferromagnetic and quantum paramagnetic
%phases, respectively. The dotted line correspond the classical solution
%$\alpha_{c}=\lambda/2$.}}{}{Figure}{\special{ language "Scientific Word";
%type "GRAPHIC";  display "USEDEF";  valid_file "T";  width 424.125pt;
%height 374.4375pt;  depth 0pt;  original-width 316.0625pt;
%original-height 260.625pt;  cropleft "0";  croptop "1";  cropright "1";
%cropbottom "0";  tempfilename 'M4E0JH00.wmf';tempfile-properties "XPR";}}}%
%BeginExpansion
\begin{figure}[ptb]
\begin{center}
\includegraphics[natheight=260.625000pt,natwidth=316.062500pt,height=374.4375pt,width=424.125pt]{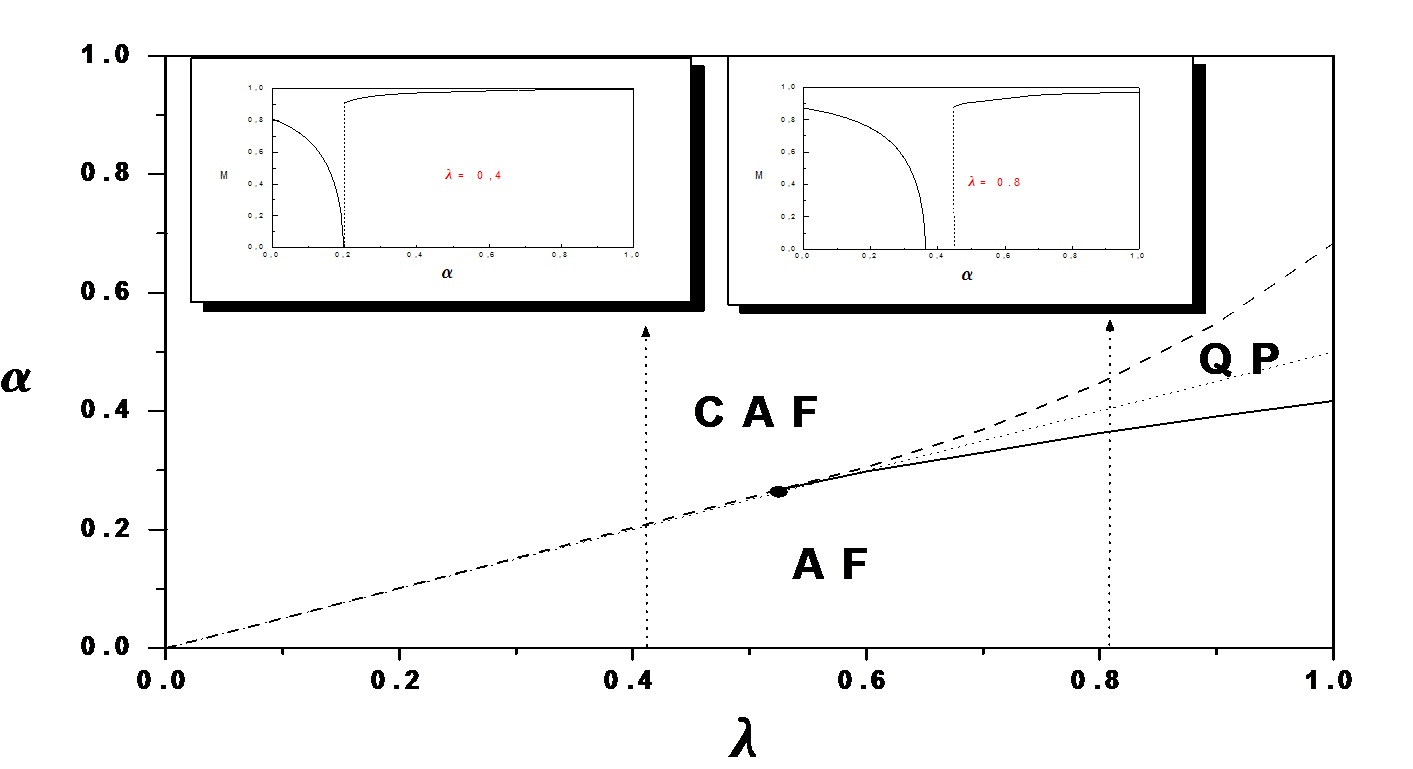}
\caption{Ground state phase diagram in the $\lambda-\alpha$ plane for the
quantum spin-$1/2$ $J_{1}-J_{1}^{\prime}-J_{2}$ model on a square lattice,
where $\alpha=J_{2}/J_{1}$ and $\lambda=J_{1}^{\prime}/J_{1}$. The dashed and
solid lines corresponds the first- and second-order transitions lines,
respectively. The black point represents the quantum critical endpoint
(\textbf{QCE}). The notations indicated by \textbf{AF}, \textbf{CAF} and
\textbf{QP} corresponds the antiferromagnetic, collinear antiferromagnetic and
quantum paramagnetic phases, respectively. The dotted line correspond the
classical solution $\alpha_{c}=\lambda/2$.}
\end{center}
\end{figure}
%EndExpansion

\section{Conclusion}

In summary, we have studied the effects of quantum fluctuations due to spatial
($\lambda$) and frustration ($\alpha$) parameter in the quantum spin-$1/2$
$J_{1}-J_{1}^{\prime}--J_{2}$ Heisenberg model. Using a variational method we
calculated the sublattice magnetization for the \textbf{AF} and \textbf{CAF}
phases. For values of $\lambda>0.51$ the frustration contributes significantly
to the existence of a disordered intermediate state (\textbf{QP}) between the
two \textbf{AF} and \textbf{CAF }ordered phases, while for $\lambda<0.51$, we
have a direct first-order transition between the \textbf{AF} and \textbf{CAF}
phases. We have observed, by analyzing the order parameters of the \textbf{AF}
and \textbf{CAF} phases, that the phase transitions are of second and
first-order between the \textbf{AF-QP} and \textbf{CAF-QP}, respectively. The
obtained phase diagram can be compared with recent results which used
effective-field theory\cite{equivalence8} and coupled-cluster
method\cite{equivalence4}, showing the same qualitative results predicting a
paramagnetic region for small interlayer parameter (i.e., $\lambda>\lambda
_{1}$), and for $\lambda<\lambda_{1}$ this \textbf{QP} phase disappears by
presenting a direct first-order transition between the \textbf{AF} and
\textbf{CAF} phases. On the other hand, recent calculations of second order
spin wave theory\cite{equivalence7} have indicated that the intermediate
\textbf{QP} phase exists for all $\lambda\leq1$ in accordance with results of
exact diagonalization\cite{12}. We speculate that by using a more
sophisticated method, for example, quantum Monte Carlo simulations\cite{mcq}
and density matrix renormalization group (DMRG) method\cite{dmrg}, this
disordered region should disappear for certain values of $\lambda<\lambda_{1}$.

\textbf{ACKNOWLEDGMENTS:} This work was partially supported by CNPq (Brazilian agency)
\end{document}